# Meta-GPT: Decoding the Metasurface Genome with Generative Artificial Intelligence

David Dang[1,2,3], Stuart Love[1], Meena Salib[1], Quynh Dang[1], Samuel Rothfarb[3,4], Mysk Alnatour[1], Andrew Salij[3], Hou-Tong Chen[2], Ho Wai (Howard) Lee[1*], Wilton J.M. Kort-Kamp[3*]

[1]*Department of Physics and Astronomy, University of California Irvine, Irvine, CA 92617, USA*

[2]*Center for Integrated Nanotechnologies, Los Alamos National Laboratory, NM 87545, USA*

[3]*Theoretical Division, Los Alamos National Laboratory, NM 87545, USA*

[4]*School of Civil and Environmental Engineering, University of Connecticut, Storrs, CT 06269, USA*

*Corresponding Authors: Howardhwlee@uci.edu, kortkamp@lanl.gov*

## Abstract

Advancing artificial intelligence for physical sciences requires representations that are both interpretable and compatible with the underlying laws of nature. We introduce METASTRINGS, a symbolic language for photonics that expresses nanostructures as textual sequences encoding materials, geometries, and lattice configurations. Analogous to molecular textual representations in chemistry, METASTRINGS provides a framework connecting human interpretability with computational design by capturing the structural hierarchy of photonic metasurfaces. Building on this representation, we develop Meta-GPT, a foundation transformer model trained on METASTRINGS and finetuned with physics-informed supervised, reinforcement, and chain-of-thought learning. Across various design tasks, the model achieves <3% mean-squared spectral error and maintains >98% syntactic validity, generating diverse metasurface prototypes whose experimentally measured optical responses match their target spectra. These results demonstrate that Meta-GPT can learn the compositional rules of light-matter interactions through METASTRINGS, laying a rigorous foundation for AI-driven photonics and representing an important step toward a metasurface genome project.

**Introduction**

For centuries, the control of light was practiced more as an art than a science. Ancient glassmakers discovered vivid structural colors in objects like the Lycurgus cup through the accidental inclusion of metallic nanoparticles,[1-3] while medieval artisans crafted lenses and mirrors by intuition and skill long before optical theory could explain them.[4] These achievements marked milestones of human creativity but lacked the systematic principles that modern optical science now provides.

That tension between artistry and scientific understanding continues in modern optics. Optical metasurfaces and metamaterials,[5, 6] arrays of subwavelength nanostructures, have transformed photonics into a platform for flat lenses,[7, 8] perfect absorbers,[9, 10] holograms,[11, 12] and quantum devices.[13] Yet, their design process is complex, with scientists needing to select from a vast combinatorial space of materials, stacks, shapes, and lattices, each coupled through resonant modes and interparticle interactions.[14] Early machine learning methods,[15] including particle swarm optimization,[16] genetic algorithms,[17] and adjoint algorithms,[18] enabled the first successes in automated metamaterial optimization. However, they relied on repeated numerical simulations rather than model-based learning from known data. A fully scalable, data-driven paradigm for metasurface design emerged with deep learning.

Deep learning and generative models have since advanced metasurface design considerably.[19] Neural networks-based surrogate solvers have yielded the quick design of ultra-high performance plasmonic and dielectric nanostructures,[20, 21] while variational autoencoders have enabled state-of-the-art beam steering metasurfaces[22] and thermal emitters.[23] Generative adversarial networks[24] and diffusion models[25] have also been applied to design terahertz structures, revealing the power of data-driven photonics. With the advent of the transformer,[26] large language models and artificial intelligence (AI) agentic systems have also emerged as a tool for designing metasurfaces, automating computational workflows from device generation to screening and evaluation.[27, 28] However, inverse design remains brittle, expensive, and fragmented as the desired properties and applications space of metasurfaces grow more complex. The central obstacle is representation: pixel maps and parameter lists lack a compositional structure that mirrors how metasurfaces are built and reasoned about, making design workflows fragile, inefficient, and difficult to scale.

Other sciences transcended similar bottlenecks when symbolic representations emerged. Chemistry advanced beyond intuition when the elements were arranged in the periodic table.[29] Biology made a similar leap with the advent of DNA sequencing, whose digital encoding through the FASTA format[30-32] (Fig. 1a) transformed genetic data into a computational language. More recently, SMILES[33] and SELFIES[34] encodings represented molecules as text, transforming chemistry into an information science[35, 36] (Fig. 1b). Once these materials were readable as sequences, they became searchable, interpretable, and ultimately, designable by both algorithms and humans. A comparable shift is now both natural and advantageous for metasurfaces.

We introduce a language-based symbolic representation that, when coupled with foundation models, establishes a systematic, AI-driven science of photonic design. In this framework, the building blocks of nanostructures are expressed as text tokens, and complete assemblies as ordered sequences, together forming a language we call METASTRINGS (METAsurface STRucture INterpretable Grammar Syntax), Fig. 1c. The language is valid by construction: every sequence obeys physical and compositional constraints and can be compiled directly for electromagnetic simulation or fabrication. Analogous to how molecular string notations reshaped chemistry, METASTRINGS provides the cornerstones for a metasurface genome project: a framework that makes photonic design feasible and interpretable at scale.

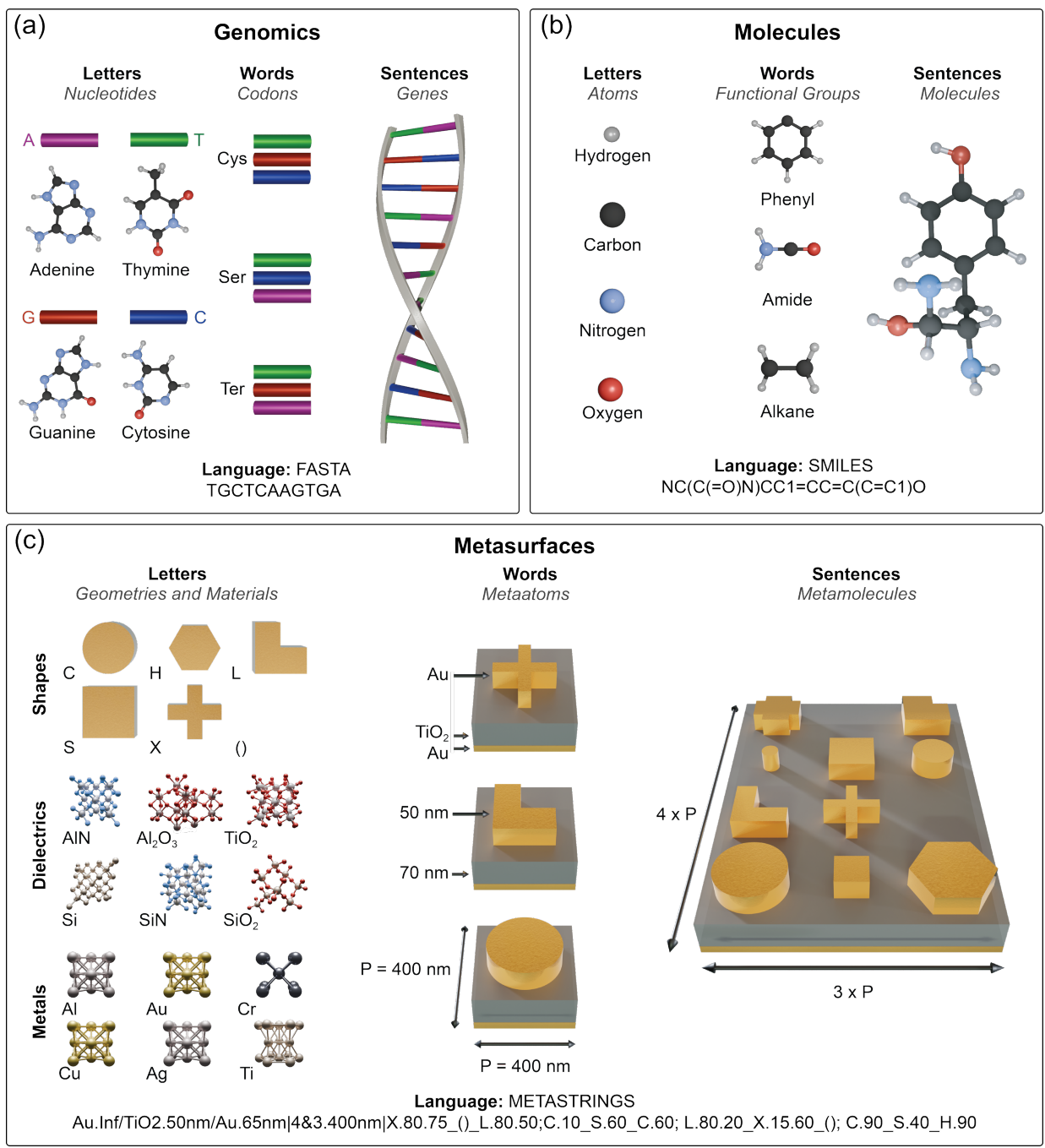

**Figure 1:** Language based representations for genomics, molecules, and photonic-metasurfaces. (a) The basic structure of DNA (deoxyribonucleic acid), with nucleotides, codons, and genes, serving as the letters, words, and sentences, respectively, in the FASTA language. (b) In chemistry, atoms, functional groups, and molecules hold a similar meaning in the SMILES strings. (c) For metasurfaces, the geometries and materials jointly function as the letters, the metaatoms as the words, and the metamolecules as the sentences in our METASTRINGS representation.

Building on this foundation, we developed a generative pretrained transformer (GPT) trained on corpora of METASTRINGS. The model, Meta-GPT, has been finetuned through supervised learning, reinforcement learning, and a chain-of-thought strategy that incorporates intermediate reasoning tokens for enhanced design accuracy. While a few works have used large language models for metasurface design,[27, 28, 37] they relied on natural language processing (i.e., long passages of text to describe a metasurface) or very simple parameter-list encodings. Although human language is effective for communication, it is not the "natural" language of metasurfaces, i.e., a representation that enables AI systems to reason more efficiently about photonic structures while remaining intelligible to humans. Meta-GPT performs exceptionally well in the inverse design of metasurfaces across diverse target spectra, generating structures that we fabricated and verified experimentally. This capability opens the door to scalable and interpretable metasurface design, establishing a new language-driven paradigm for photonic science. Beyond performance, the success of Meta-GPT signals a broader era of generative AI-driven scientific discovery through the development of domain-specific linguistic representations for other physical sciences.

**Results**

*METASTRINGS Representation*

METASTRINGS is a symbolic representation that encodes the structural hierarchy of metasurfaces as an ordered sequence of text tokens, as shown in Fig. 1c. Each valid sequence corresponds to a realizable design, enabling systematic exploration of photonic architectures through linguistic generation rather than geometric parameterization. Every METASTRINGS consists of three segments, separated by vertical bars "|" in the syntax, that mirror the physical organization of the device:

1. **Material stack** – Defines the ordered layers of the metasurface, where each layer is written as "$material.thickness$" and layers are divided by "/".
2. **Lattice configuration** – Specifies the number of rows and columns of the metamolecule separated by an ampersand "&", followed by the metaatom period.
3. **Geometrical pattern** – Lists the shapes patterned on the top layer, each denoted by a letter followed by one or two integer numbers approximating its dimensions as a percentage of the metaatom cell size. Underscores "_" separate shapes within the same row, and semicolons ";" separate successive rows.

The METASTRINGS vocabulary includes material identifiers, geometric primitives, numeric parameters, and grammatical delimiters. Together these form the language's dictionary, defining the symbols that can be combined into valid sequences. For clarity, we consider here (Fig. 1c) a subset of the full possible vocabulary that captures key structural relationships within experimentally accessible designs. However, the vocabulary can be expanded to accommodate new design objectives and emerging fabrication capabilities. This flexibility ensures that the language remains adaptable to future photonic platforms.

The structure and meaning of the METASTRINGS are illustrated by the example in Fig. 1c:

$$Au.Inf/SiN.50nm/Au.65nm$$
$$|4\&3.400nm|$$
$$X.80.75\_()\_L.80.50; C.10\_S.60\_C.60; L.80.20\_X.15.60\_(); C.90\_S.40\_H.90$$

This sequence encodes an Au-SiN-Au metal-dielectric-metal stack patterned into a 4 × 3 array of metaatoms, each with a 400 nm unit cell period. The first segment (red) specifies the layer compositions and thicknesses (where "Inf" represents an effectively infinite thickness for the substrate layer in simulations), the second (green) defines the lattice geometry, and the third (blue) lists the shapes on the patterned layer. For example, the sequence "X. 80.75_()_L. 80.50" defines three metaatoms: a cross with arms 80% wide and 75% high relative to the cell size, an empty site (), and an L-shape with 80% width and 50% length. The remaining rows follow the same syntax to combine circles, squares, hexagons, and L-shapes in different proportions.

The METASTRINGS syntax maintains a consistent hierarchy in which the material stack precedes the lattice and geometry segments. The delimiters "/", "|", "&", "_", "nm", and ";" appear in fixed order, enforcing consistent sequences that can be parsed unambiguously for automatic validation and ensuring that any string conforming to the syntax can be compiled into a three-dimensional geometry. At the semantic level, each token corresponds to a particular design feature: material symbols encode optical constants, numerical parameters specify dimensions, and delimiters describe spatial relationships within the lattice (see Methods for the full list of parameters). Through this mapping, METASTRINGS provides a deterministic connection between text and structure, allowing straightforward translation into models for simulation and fabrication.

A notable feature of METASTRINGS is its ability to reproduce familiar metasurface configurations as limiting cases within the same syntax. For instance, a standard Bragg mirror is recovered when the geometry segment contains only empty cells, represented by "()", which removes the top metaatom layer and yields an unpatterned multilayer stack. A single metaatom device is obtained with a 1 × 1 lattice. More generally, periodic metamolecule arrays are described naturally through N × M lattices, providing a compact symbolic counterpart to conventional unit cell parameterizations. These examples show that METASTRINGS subsumes traditional photonic design representations within a unified grammar. Unlike parameterized models that restrict design to fixed shapes or feature lists,[24] METASTRINGS defines a generative space that grows with the vocabulary itself. Whereas pixel representations[38] are powerful for dense, subwavelength local pattern optimization, our symbolic language formulation captures compositional hierarchy, enabling reasoning about materials, lattice structure, and geometry simultaneously. Together, these features elevate metasurface design from brute-force numerical search to a language-driven exploration of structure and function.

*Development of a Foundation Model for Metasurfaces*

Building on the METASTRINGS representation, we demonstrate that this language can be learned autonomously by a large language model tailored for photonic design. To this end, we developed Meta-GPT, a foundation transformer pretrained exclusively on the METASTRINGS corpus and implemented from the ground up (full hyperparameters, model details, training efficiency, and computational setup are provided in the Methods and Supplementary Information). The objective of this stage is to enable the model to internalize the grammatical and compositional structure of the language without relying on any simulation or experimental data, forming the basis for downstream physics-informed finetuning.

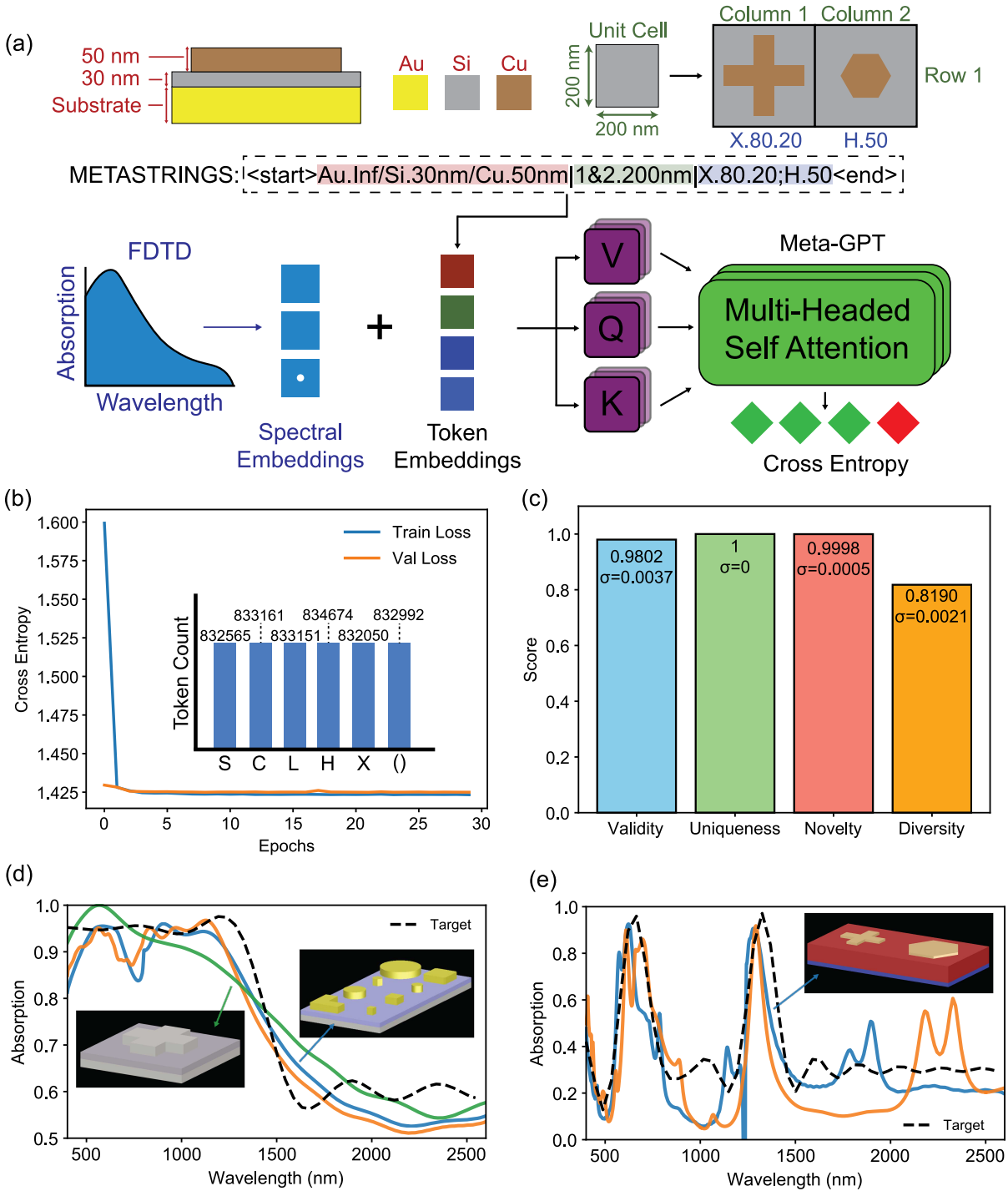

**Figure 2:** Inverse design with Meta-GPT. (a) Schematic of the pretraining and supervised finetuning for the Meta-GPT foundation model. During pretraining, the model is trained only on METASTRINGS to learn the grammatical structure of the metasurface genome through next-token prediction. Finetuning conditions the METASTRINGS on simulated absorption spectra, enabling zero-shot inverse design. (b) Pretraining loss history showing model convergence within 5 epochs. (c) Statistical performance metrics of Meta-GPT, including validity, uniqueness, novelty, and diversity of newly generated METASTRINGS. Examples of inversely designed metasurfaces proposed by the finetuned Meta-GPT and their simulated absorption (solid curves) for (d) short-wavelength and (e) dual-band absorber targets (dashed).

For this demonstration, we restricted the language to metal-dielectric-metal (MDM) metasurfaces, a widely studied architecture that captures the essential physics of resonant absorption while remaining experimentally accessible. The parameter space was defined from a literature survey of experimentally verified MDM metasurfaces operating in the visible to infrared range[39-43]. The design space for METASTRINGS generation is described in Methods. The vocabulary, comprising six metallic materials, six dielectrics, and five geometric primitives, is summarized in Fig. 1c.

Figure 2a schematically illustrates the training workflow: each metasurface design is encoded as METASTRINGS and provided as input to the transformer model, which learns token relationships through next-token prediction within the language's lexicon. The pretraining dataset consisted of one million syntactically-valid METASTRINGS generated randomly, ensuring balanced coverage of possible geometries and material combinations across the design space (see inset to Fig. 2b). Each sequence was tokenized semantically, with material thickness, delimiters, and shape-parameter pairs treated as distinct tokens to preserve the language hierarchy. Spectral data, shown schematically in Fig. 2a, are concatenated to the left of the text sequence during fine-tuning but omitted during pretraining. The model converged rapidly, with the loss stabilizing after five epochs (Fig. 2b), confirming that it successfully learned the METASTRINGS grammar.

To assess the generative capacity of the pretrained model, we evaluated the validity, uniqueness, novelty, and diversity of the sequences produced by Meta-GPT, as summarized in Fig. 2c. A total of 50,000 METASTRINGS were sampled across 50 trials, yielding 98% valid outputs, i.e., sequences that followed the correct syntax and grammatical constraints. All generated sequences were unique, and more than 99% were novel relative to the training corpus. The diversity of generated designs, quantified using a Lichtenstein-like distance metric (Methods), reached ~82%, confirming that the model explored a broad region of the design space. These results demonstrate that Meta-GPT learned the syntactic and combinatorial structure of the METASTRINGS and can generate physically meaningful and compositionally diverse metasurfaces.

*Supervised Finetuning and Inverse Design*

After pretraining, Meta-GPT was finetuned to perform inverse metasurface design tasks using spectral data as conditioning input. A dataset of 20,000 METASTRINGS and their corresponding absorption spectra, simulated with Lumerical finite-difference time-domain (FDTD) solvers,[44] was used for this stage. During finetuning, each spectrum was concatenated with its associated METASTRINGS, enabling the model to learn correlations between spectral responses and structural encodings. Once trained, the model accepts a target absorption spectrum as input and outputs METASTRINGS that can be compiled into geometries for simulation or fabrication.

We evaluated the finetuned model on two representative metasurface design scenarios, namely short-wavelength and dual-band absorbers, selected to probe distinct spectral behaviors of MDM structures. Figures 2d,e shows the target spectra together with the responses of the three highest-scoring generated metasurfaces. Insets display corresponding metasurface designs from each case, illustrating the diversity of geometries produced by the model. Across both scenarios, the simulated absorption spectra of the generated designs closely match the targets, achieving mean-squared-error values below 0.03 across the visible to infrared spectral range.

Beyond quantitative agreement, the generated sequences exhibit varied combinations of materials, geometric primitives, and lattice configurations (see Supplementary Information for a full list of inverse-designed METASTRINGS and their simulated spectra), demonstrating that Meta-GPT retained the compositional diversity learned during pretraining while adapting to spectral conditioning. This diversity enables the identification of designs compatible with practical synthesis and fabrication constraints, including material availability, aspect-ratio limits, deposition control, and lithographic resolution, ensuring that generated structures remain experimentally viable. These results confirm that a language-trained transformer can successfully transfer from symbolic pretraining to physics-aware photonic design, establishing a robust baseline for subsequent optimization via more advanced training approaches.

*Physics-Informed Reinforcement Learning*

To further improve the baseline finetuned model's performance, we implemented a reinforcement learning (RL) strategy inspired by physics-informed and RL-based design frameworks such as PHORCED[45] and GLOnet.[46, 47] Figure 3a schematically summarizes the workflow, in which the model receives a target absorption spectrum and iteratively refines its generation policy through simulated feedback. The model parameters were initialized from the supervised baseline and updated using the REINFORCE policy-gradient algorithm.[48] The reward was defined as the negative mean-squared error (MSE) between the target and simulated spectra, weighted by the log-probability of generated METASTRINGS. We trained the RL model for 15 epochs using the full FDTD solver to provide accurate rewards; during each epoch, 30 METASTRINGS were sampled to compute average rewards, ensuring stable optimization while maintaining computational efficiency.

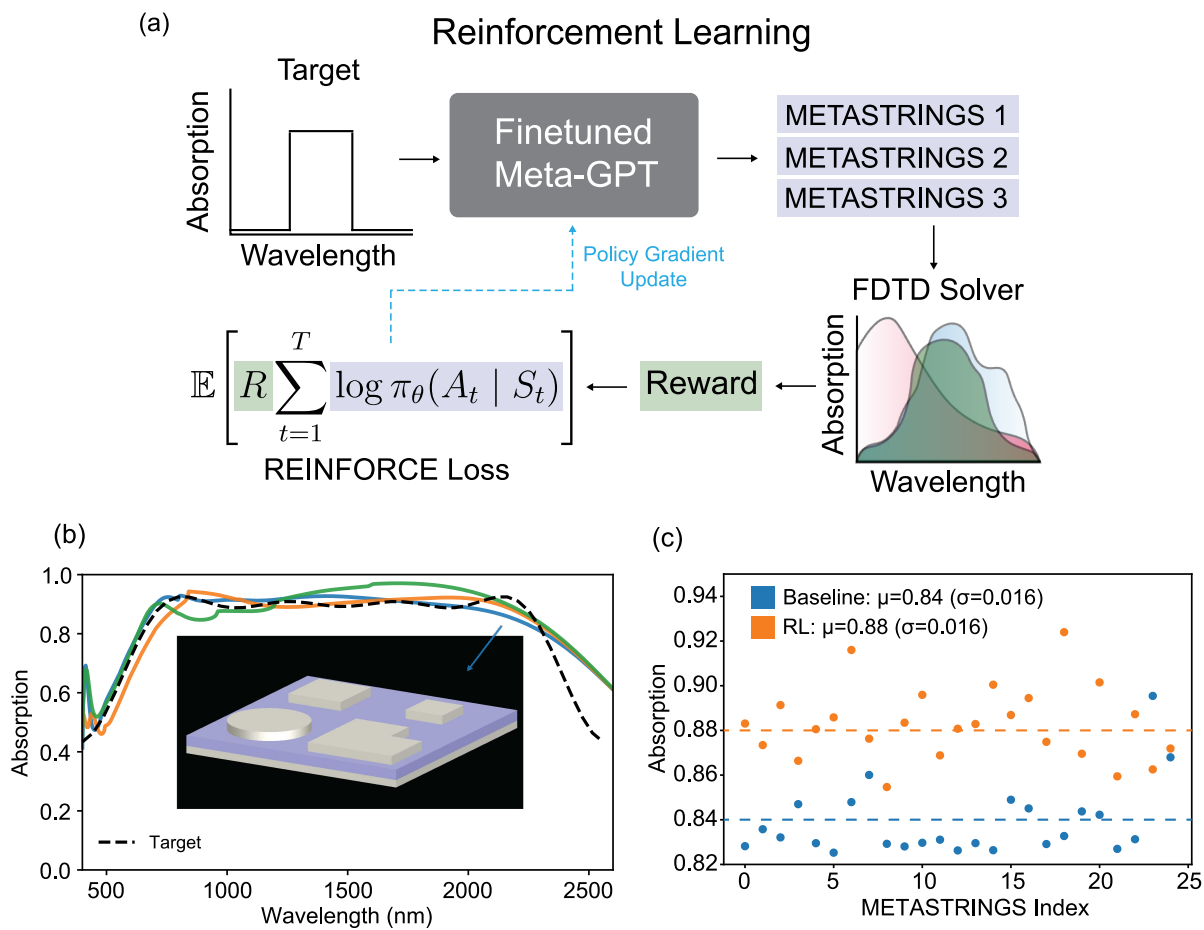

**Figure 3:** Reinforcement learning with Meta-GPT. (a) The REINFORCE algorithm employs the negative mean-squared error between the target spectrum and the simulated one obtained via a FDTD solver as the reward for Meta-GPT. (b) Absorption spectrum (solid curves) for examples of RL-generated metsurfaces for a broadband absorber target (dashed). (c) Average absorption of the top 20 metasurfaces generated with the baseline and RL models. We calculate the average absorption between 700 nm and 2300 nm. The orange line represents the average absorption (μ) at 0.88 for RL compared to 0.84 for baseline (blue line).

The RL-trained Meta-GPT generates ultra-broadband absorbers whose simulated spectra closely reproduce the target response, as shown in Fig. 3b for the three best-performing designs. A broader statistical comparison is presented in Fig. 3c, where each point corresponds to the average absorption for each of the top twenty generated metasurfaces. The RL model clearly shifts toward higher absorption values relative to the finetuned baseline, indicating progressive improvement as physical feedback is incorporated. This gain arises from the model's enhanced ability to refine geometrical features and material selections to strengthen local resonances, the dominant absorption mechanism in MDM structures. These findings confirm that RL effectively couples symbolic generation with physical feedback, enabling autonomous exploration and optimization of design parameters within the METASTRINGS framework. While all results shown here employ FDTD-based reward functions, we also introduce a surrogate-assisted RL variant (see Supplementary Information) that uses a deep neural network to simulate the absorption spectra of the METASTRINGS. The resulting performance and accuracy lies between that of the finetuned and FDTD-based RL models, while enabling a nearly 300x speedup compared to computationally costly, full-wave simulations.

*Chain-of-Thought in METASTRINGS*

Chain-of-thought (CoT) prompting[49] is a relatively simple yet powerful strategy that enhances a large language model's ability to reason through multi-step problems. By exposing a model to explicit sequences of intermediate logical steps, CoT training significantly improves success rates on tasks that demand structured logic (Fig. 4a, top). Here, we extend this concept to photonic design, investigating how CoT reasoning within the METASTRINGS framework influences the generated absorption spectra relative to our baseline model. To our knowledge, this represents the first demonstration of CoT reasoning applied to photonic design.

In our implementation, we augmented the training and validation datasets with CoT tokens placed before those that define a standard METASTRINGS (Fig. 4a, bottom). These CoT tokens consist of two parts: (1) the materials for each layer and (2) the geometric features, specifying shape and relative dimensions. To encourage reasoning rather than direct copying, the thickness, period, and lattice dimensions were intentionally withheld. The intuition is that the model first identifies which materials and shapes correspond to the target spectrum through the CoT and then infers missing parameters. Random dropout was applied during training to promote generalization between the CoT sequence and final METASTRINGS.

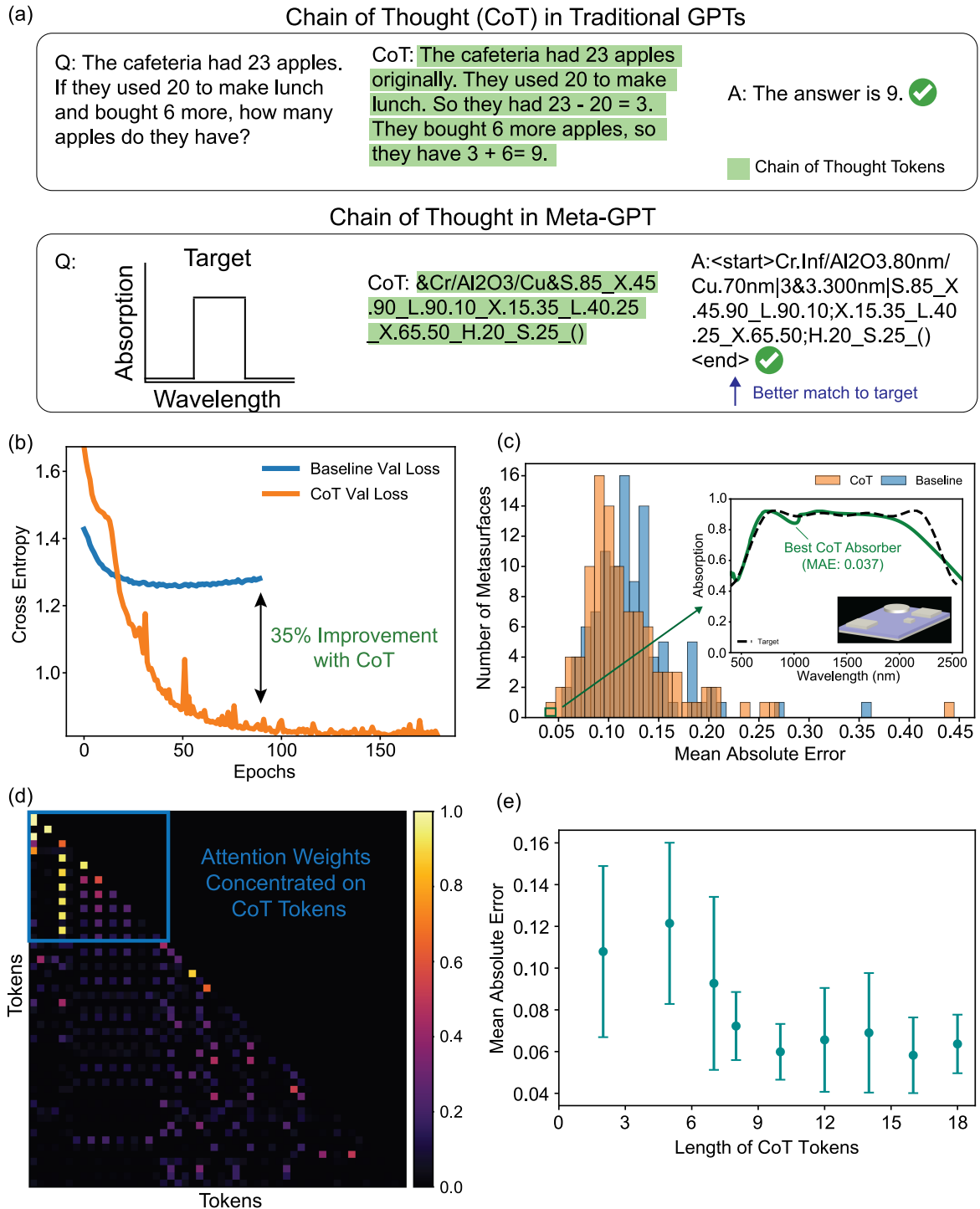

**Figure 4:** Chain-of-thought reasoning in Meta-GPT. (a) Comparison of CoT approaches in a typical large language model[49] (top) and our CoT implementation in METASTRINGS (bottom). (b) Validation loss during supervised training for baseline and CoT models. (c) MAE distribution between a target broadband absorber spectrum and that of 100 metasurfaces generated by the baseline and CoT models. The inset shows the best performing absorber CoT design (green) closely matching the target (dashed). (d) Attention map from the final transformer layer highlights strong focus on CoT tokens for the lowest MAE CoT broadband absorber (see Methods). (e) Performance of generated METASTRINGS with limited CoT tokens.

During training, the CoT-enhanced Meta-GPT achieved a ~35% reduction in validation cross-entropy loss compared to the baseline model (Fig. 4b). To assess performance, we tasked the CoT model with generating one hundred samples for a broadband absorber operating between 400 nm and 2600 nm (see inset in Fig. 4c). The resulting spectra exhibited markedly lower mean absolute error (MAE) values relative to the baseline, as shown in the histograms in Fig. 4c. Moreover, the top-performing CoT-generated absorber achieved a 29.7% reduction in MAE relative to the best baseline design, nearly reproducing the target absorption spectrum between 400 nm and 2000 nm.

To gain a deeper insight into how the model leverages the CoT information, we analyzed the attention weights from the final transformer layer across multiple generated samples. As shown in Fig. 4d, the model consistently assigns high attention values to the CoT tokens. This indicates that the model actively uses the CoT sequence as a reasoning scaffold to infer missing geometric and structural parameters when inverse-designing a metasurface that matches the target spectrum.

During training, the model was exposed to a mixture of full and truncated CoT sequences, where each shortened variant omitted part of the reasoning chain (see Methods for examples). Figure 4e shows the MAE for the broadband absorber of Fig. 4c as a function of CoT-token length. Each point represents the average over ten generated designs. Performance improves steadily with longer CoT sequences, reaching an optimum region around a token length of ten, corresponding to the minimum MAE and smallest standard deviation. Beyond this point, both the MAE and its variation increase slightly, suggesting that excessively long reasoning chains may introduce redundant or noisy information. These results indicate that CoT tokens provide physically meaningful context that guides the generation process, enabling Meta-GPT to reason more effectively about metasurface design within the METASTRINGS framework.

**Discussion**

Table 1 compares the performance of the baseline, RL, and CoT models on the broadband absorber target, quantified through mean-squared error calculated from one hundred generated and simulated designs. Both RL and CoT models outperform the baseline, producing more accurate and reliable results. In contrast, the RL model achieves the highest stability and accuracy through physics-informed feedback learned during training. The CoT model performs similarly well on average and often surpasses the RL model. However, its results show higher variance, reflecting its broader exploratory reasoning process over the design space. Overall, the RL approach offers optimization efficiency and consistency, while the CoT framework enhances diversity and interpretability.

| Model | Median | Min | Max | Mean | Std. |
|---|---|---|---|---|---|
| Baseline (MSE) | 0.0226 | 0.0038 | 0.158 | 0.0259 | 0.02 |
| RL (MSE) | **0.0176** | 0.0036 | **0.0815** | **0.0201** | 0.0129 |
| CoT (MSE) | 0.0185 | **0.0031** | 0.2251 | 0.0264 | 0.0271 |

**Table 1:** Mean-squared error comparison between baseline, RL, and CoT models on the broadband absorber target. Each model generated one hundred METASTRINGS and the absorption spectra was validated via FDTD solvers.

These results demonstrate that a compact, text-based representation of metasurfaces through METASTRINGS enables large language models to internalize and exploit the underlying physical design principles of these structures. As shown with Meta-GPT, the model can generate metasurfaces that accurately reproduce target absorption spectra. The finetuned baseline model establishes a foundation for zero-shot inverse design, while the RL variant enhances performance consistency and overall absorption accuracy through iterative physics-informed feedback. Complementarily, the CoT model introduces a reasoning layer that connects physical relationships to design choices. Collectively, these strategies move Meta-GPT beyond pattern recognition toward physics-aware, interpretable metasurface generation, providing an autonomous framework that can reason about light-matter interactions through symbolic design rules.

Looking ahead, the combination of METASTRINGS and Meta-GPT offers a scalable route toward language-based design automation in photonics. Here, we demonstrated METASTRINGS for MDM nano-photonic architectures using a finite set of shapes and materials, but the framework can be easily extended to encompass a broader range of photonic systems. The same syntax can describe phase-gradient metasurfaces for beam steering,[50] e.g., through an N × 1 array of rectangular elements defined as "$R.length.width.rotation angle$," or holographic and imaging devices where phase control is the dominant design goal.[51,52] Additional patterned layers or polarization descriptors can be incorporated within the same grammar, providing a path toward a general photonic language capable of describing diverse metasurface functionalities. Finally, coupling this symbolic representation with multimodal data, including spatial field distributions from simulations, experimental spectra, and fabrication metadata, could enable Meta-GPT to reason across the full design-fabrication-characterization pipeline. Beyond photonics, METASTRINGS can be generalized to other physical domains, including mechanical,[53,54] thermal,[55] and quantum[56,57] metasurfaces and metamaterials, unifying their design through shared principles of wave propagation and energy confinement. This direction points toward a generalized language-based inverse-design paradigm for multiphysics systems, where structure, material, and field interactions are jointly optimized through generative reasoning.

**Methods**

*Python to Electromagnetic Field Solver Parsing*

The parser, implemented in Python, reads each METASTRINGS from left to right, reconstructs the structure row by row, and automatically generates the input geometry for finite-difference time-domain simulations (*Ansys Lumerical*). Because all tokens are deterministic, a string can be compiled directly into a three-dimensional model or fabrication layout without manual intervention. This capability allows symbolic sequences—whether human-written or generated by language models—to be simulated and validated immediately.

*Model Pretraining and Finetuning*

Our model is a 12-layer neural network, with 12 attention heads in each layer – mimicking the design of GPT-2. The total vocabulary size is 694, with a token embedding dimension of 5. The spectral embedding dimension is 50 (~44 nm resolution in the spectrum) and the total context size for the transformer stack is 140.

All models used an 80/20 split between training and validation sets as well as a learning rate of 0.0003 and weight decay of 0.001, using an AdamW optimizer. All trainings used model checkpointing, keeping the lowest average cross entropy on the validation dataset after every epoch. We performed the pretraining process on Los Alamos National Laboratory's supercomputer cluster Chicoma, which used 4 A100 GPUs to train our model on 1,000,000 METASTRINGS (without physical simulations) in roughly 40 minutes for 30 epochs.

For the finetuning process of the baseline and CoT model, we used a local RTX 3090 with an Intel 12th Gen Core™ i5-12600K, resulting in training times of 26 minutes and 43 minutes, respectively, for 90 epochs. Both models used a batch size of 32 for both the training and validation and an overall dataset of 20,000 METASTRINGS with simulated absorption spectra.

The CoT-models were trained on our previous dataset of METASTRINGS with appended CoT tokens at the beginning of each METASTRINGS. We used the foundation Meta-GPT model as the starting point for the CoT model training and trained for 180 epochs. To allow the model to understand partial CoT tokens for Figure 4e, we randomly truncated the total length of the CoT tokens, still keeping the total number of METASTRINGS at 20,000, and trained the model using this dataset for 180 epochs.

The RL model training used a separate computer with an RTX 3060 and Intel 13th Gen Core ™ i5-13600k, resulting in 55 hours for 15 epochs due to the slow Lumerical FDTD solver. Note that we used the baseline model as the starting point for the RL model training instead of starting from the foundation model. Each epoch sampled 30 METASTRINGS to ensure accurate rewards averages while also balancing computational time.

*Lumerical FDTD Simulations*

For Lumerical FDTD simulations, we used UC Irvine's Supercomputing Cluster, Greenplanet in parallel on ~50 nodes at maximum capacity. We were typically able to conduct 10,000 Lumerical FDTD at medium mesh size (10 mesh layers per cell and ~10 nm mesh in x- and y- directions) over 10-14 days.

Our METASTRINGS were simulated with a selected set of metals and dielectrics for the MDM stack, using their refractive index data over the 400 nm – 2600 nm. The listed metals[58] include: Al, Au, Cr, Cu, Ag, and Ti. For dielectrics: AlN,[59] $Al_2O_3$,[58] $TiO_2$,[60] Si,[58] SiN,[61] and $SiO_2$.[58] The design space encoded by the METASTRINGS spans dielectric thicknesses from 40 nm to 250 nm (in 50 nm increments) and metallic layer heights from 20 nm to 70 nm (in 5 nm increments). The meta-atom period ranges from 200 nm to 600 nm, and the metasurface layout varies between 1 × 1 and 4 × 4 arrays.

*Diversity Metric with Levenshtein Distance*

For the diversity metric in Fig. 2b, we defined a Levenshtein-like distance metric by removing redundant delimiter tokens, such as "/", "|", "&", "_", "nm", and ";". This metric is given by

$$D = \frac{2}{N(N-1)} \sum_{i<j} \frac{\text{dist}(S_i, S_j)}{\max(S_i, S_j)}.$$

Here, N is the total number of METASTRINGS generated in a batch, $\text{dist}(S_i, S_j)$ is the Levenshtein distance at the token level between METASTRINGS $S_i$ and string $S_j$, and $\max(S_i, S_j)$ is the largest token length between the pair of METASTRINGS. Note that this definition enforces that *D* is bound to the 0 to 1 range.

*Listed METASTRINGS*

This section lists the METASTRINGS for all absorption plots shown in this work:

**Fig. 2d:**
Blue curve:
Ti.Inf/SiN.40nm/Au.60nm|4&2.200nm|C.90_X.50.55;H.25_C.25;C.50_S.30;L.70.30_S.15

Orange curve:
Ti.Inf/Al2O3.40nm/Ag.65nm|2&4.200nm|L.55.15_S.35_L.25.90_H.45;C.15_H.90_X.20.10_L.40.45

Green curve:
Ti.Inf/Al2O3.40nm/Ti.60nm|1&1.350nm|X.25.60

**Fig. 2e:**
Blue curve:
Al.Inf/AlN.210nm/Cu.30nm|1&2.500nm|X.15.55_H.70

Orange curve:
Al.Inf/AlN.220nm/Al.20nm|2&1.600nm|S.55;X.65.60

**Fig. 3b:**
Blue curve:
Cr.Inf/SiN.70nm/Cr.60nm|2&2.400nm|S.50_S.35;C.70_L.50.80

Orange curve:
Cr.Inf/AlN.70nm/Cr.65nm|3&4.350nm|C.90_S.20_X.35.90_L.45.20;S.10_H.15_L.55.80_S.80;S.50_C.90_L.45.10_L.35.30

Green curve:
Cr.Inf/Al2O3.80nm/Cr.70nm|3&4.400nm|H.75_S.60_H.90_C.90;X.30.65_S.65_L.10.70_S.35;X.45.65_L.20.45_H.45_L.80.55

**Fig. 4c:**
Green curve:
CoT: &Cr/SiN/Cr&C.75_S.65_()_S.25_L.65.70_()
Cr.Inf/SiN.70nm/Cr.70nm|3&2.350nm|C.75_S.65;()_S.25;L.65.70_()

*Partial CoT Tokens*

For Fig. 4e, we used shortened CoT tokens as the input into our model:

&Cr

&Cr/SiN/
&Cr/SiN/Cr&
&Cr/SiN/Cr&C.75
&Cr/SiN/Cr&C.75_S.65
&Cr/SiN/Cr&C.75_S.65_()
&Cr/SiN/Cr&C.75_S.65_()_S.25
&Cr/SiN/Cr&C.75_S.65_()_S.25_L.65.70
&Cr/SiN/Cr&C.75_S.65_()_S.25_L.65.70_()

**Acknowledgements**

DD acknowledges support by the UCI-LANL-SoCal Hub and the DOE SCGSR graduate fellowship program. SR acknowledges financial support through the UConn's Pratt & Whiteny Institute for Advanced Systems Engineering Graduate Fellowship which enabled him to contribute to this work. WJMKK and HTC acknowledge the LANL Laboratory Directed Research and Development (LDRD) program for funding under project 20250492ER. WJMKK also thanks LDRD project 20260116ER, and WJMKK and AS acknowledge LDRD project 20250006DR. We thank Dr. Nathan Crawford for assistance with the Greenplanet Supercomputing Cluster at UC Irvine as well as acknowledge NSF Grant CHE-0840513 for funding several nodes and hardware upgrades for the supercomputing cluster. This research also used resources provided by the Los Alamos National Laboratory Institutional Computing Program, which is supported by the U.S. Department of Energy National Nuclear Security Administration under Contract No. 89233218CNA000001. Part of this work was also performed at the Center for Integrated Nanotechnologies, a U.S. Department of Energy, Office of Basic Energy Sciences user facility. LANL is operated by Triad National Security, LLC, for the National Nuclear Security Administration of the U.S. Department of Energy.

**Contributions**

WJMKK conceived the idea of METASTRINGS and Meta-GPT, while DD and MS refined and enhanced it. DD created the model and performed the model pretraining, finetuning, and evaluation under guidance from WJMKK. DD also conceived the reinforcement learning and chain-of-thought approaches. MS and DD wrote Lumerical scripts to interface with the python Lumerical API for automatic translation of METASTRINGS into FDTD simulations. SL, QD, DD, and MA conducted fabrication and measured absorption spectra for selected structures proposed by the generative models under guidance from HWHL. SR, AS, and HTC provided scientific feedback regarding the development of the project. WJMKK and DD wrote the manuscript. All authors contributed to discussions, provided feedback, and approved the final version of the manuscript. HWHL and WJMKK oversaw the overall execution of the project.